%% file: third_draft.tex
\documentclass[10tpt]{elsarticle}
\usepackage{graphicx}
\usepackage{amssymb}
\usepackage{fullpage}
\usepackage{algorithm,algorithmic}
\usepackage{microtype}
\usepackage{epstopdf}
\usepackage[english]{babel}

\usepackage{natbib}

\input{macros.tex}

\title{A weakly continuous post-processing technique for stabilizing the pressure projection operator in marginally-resolved incompressible inviscid flow} 
\title{A post-processing technique for stabilizing the discontinuous pressure projection operator in marginally-resolved incompressible inviscid flow} 
\author{Sumedh M.~Joshi\footnote{Corresponding author.}}
\address{Center for Applied Mathematics, \\ 657 Rhodes Hall, Cornell University, \\ Ithaca NY, 14850 \\ smj96@cornell.edu \\ 7135945390}

\author{Peter J.~Diamessis}
\address{School of Civil and Environmental Engineering, \\ 105 Hollister Hall, Cornell University, \\ Ithaca NY, 14850}

\author{Derek T. Steinmoeller}
\address{ Department of Applied Mathematics \\ University of Waterloo \\ Ontario, CA}

\author{Marek Stastna}
\address{ Department of Applied Mathematics \\ University of Waterloo \\ Ontario, CA}

\author{Greg N. Thomsen }
\address{ Applied Research Laboratories \\ The University of Texas at Austin \\ Austin, TX }

\begin{document}
\begin{abstract}

A method for post-processing the velocity after a pressure projection is developed that helps to maintain stability in an under-resolved, inviscid, discontinuous element-based simulation for use in environmental fluid mechanics process studies.  The post-processing method is needed because of spurious divergence growth at element interfaces due to the discontinuous nature of the discretization used.  This spurious divergence eventually leads to a numerical instability.  Previous work has shown that a discontinuous element-local projection onto the space of divergence-free basis functions is capable of stabilizing the projection method, but the discontinuity inherent in this technique may lead to instability in under-resolved simulations.  By enforcing inter-element discontinuity and requiring a divergence-free result in the weak sense only, a new post-processing technique is developed that simultaneously improves smoothness and reduces divergence in the pressure-projected velocity field at the same time.  When compared against a non-post-processed velocity field, the post-processed velocity field remains stable far longer and exhibits better smoothness and conservation properties. 
\end{abstract}

\maketitle

\section{Introduction}

Pressure-projection methods are a class of numerical techniques that decouple the solution of pressure from velocity in the numerical simulation of an incompressible flow.  These methods, first developed by Chorin \cite{Chorin1968}, overcome the difficult problem of the pressure-velocity coupling through the incompressibility constraint, and are widely used in computational fluid dynamics for solving time-dependent flow \cite{Guermond1996, Guermond2006}.  The fundamental idea in the pressure-projection method is that the momentum equation and the incompressibility constraint are time-integrated separately and in sequence.  First the momentum equations are solved to advance the velocity to an interstitial time.  Next, a projection operation maps these velocities onto the space of divergence-free functions by way of solving a Poisson equation and advances the velocity to the next time.  Thus, projection methods are also sometimes known as fractional-step or time-splitting methods \cite{Karniadakis1991,Kim1985,Marcus1984} and are widely-used solving viscous, time-dependent, incompressible flows.  Because of their ubiquity, much work has been to done to construct consistent boundary conditions for the pressure and velocity \cite{Kim1985, Karniadakis1991, Sani2006, Brown2001,Liu2010, Tomboulides1989} and stable spatial discretizations of each constituent operator in the time-splitting \cite{Deville2002}.  These efforts are largely focused on avoiding the spurious divergence boundary layers that can form when inconsistent boundary conditions are used in the time-splitting within the projection method \cite{Karniadakis1991, Tomboulides1989}.   A concise review of pressure projection methods can be found in Ref. \cite{Guermond2006}.
 
While in theory the velocity, once projected, is supposed to be divergence-free, it has recently been observed that in the discontinuous Galerkin (DG) formulation the projection operation may contain non-solenoidal eigenmodes \cite{Steinmoeller2013}.  Thus, the projected velocity fields themselves are not exactly divergence-free, which can lead to numerical instability and inaccuracy .  A possible explanation for the cause  (which is discussed in section \ref{no_post_processing}) of this is that this is due to the discontinuous spatial discretization of the Laplacian operator within the pressure projection method.  To remedy this, Steinmoeller et al. \cite{Steinmoeller2013} construct a post-processing method that explicitly projects the velocity field onto a computed basis set that is exactly the null-space basis of the discrete DG divergence operator, and show that this projection eliminates the spurious divergence due to the non-solenoidal eigenmodes of the pressure projection update operator.   While this post-processing technique is effective in eliminating non-solenoidal components of the velocity, by virtue of the discontinuous nature of the spatial discretization, the exact null-space projection (ENP) as in Steinmoeller et al. does not take into account continuity between elements.  ENP is a projection that is entirely local to an element. 

In this work, we ask whether this locality is ever problematic, and if so what should be done to address it.  While this question is broad in scope, it is shown that at least in one instance of a marginally resolved simulation the discontinuity in the ENP can lead to instability.  As a remedy, a modified null-space projection technique is used as a post-processing method that explicitly takes into account inter-element continuity in a regularized least-squares sense.  It is shown that in this instance the so-called weak null-space projection method appears to yield greater stability as a post-processing technique.  In this regard, this line of reasoning is in parallel to previous efforts which focused on constructing boundary conditions to avoid the spurious divergence boundary layers \cite{Karniadakis1991, Tomboulides1989}, but focusing instead on the spurious divergence that is observed to form at inter-element boundaries.  
 
To study the relative merits of the exact null-space projection and its modification that captures inter-element continuity, the 2D incompressible Euler equations are used as a proxy for the full incompressible Navier-Stokes equations.  The incompressible Euler equations model an inviscid incompressible flow with a stratified background density profile that is not dependent on time and only depends on the vertical direction; a perturbation density, $\rho'(\bf{x},t)$, is overlaid on the background stratification and as noted does vary in time and space.  The equations are given as 

\begin{align}
	\label{euler1}
	\prtl{\bf{u}}{t} &= \bf{u} \cdot \nabla \bf{u} - \frac{1}{\rho_0} \nabla p- g \frac{\rho'}{{\rho_0}}\bf{e}_z \\
	\nabla \cdot \bf{u} &= 0 \label{euler2} \\
	\prtl{\rho'}{t} &= -\nabla \cdot \bf{u} ( \rho' + \overline{\rho} ) 
\end{align}
where $\rho(x,z,t) = \rho_0 + \overline{\rho}(z) + \rho'(x,z,t)$ is density stratified in the Boussinesq approximation with $\overline{\rho}(z)$ the background stratification, $\bf{u}(x,z,t)$ the velocity, $p$ the pressure, and $\bf{e}_z$ the unit vector in the vertical direction.  In the pressure-projection method Eqs.~(\ref{euler1})-(\ref{euler2}) are not solved directly.  Instead, a projection operator $\mathbb{P}$ is used to solve 
\begin{align}
	\label{eulerproj}
	\prtl{\bf{u}}{t} &= \mathbb{P}\left( \bf{u} \cdot \nabla \bf{u} - g \frac{\rho'}{{\rho_0}}\bf{e}_z \right) 
\end{align}
in which the projection operator $\mathbb{P}$ is defined as 
\begin{align}
\label{projection}
\mathbb{P\bf{u}} := \bf{u} - \nabla \Delta^{-1} (\nabla \cdot \bf{u}).
\end{align}
This decouples the solution of pressure from velocity, and allows for a sequential solution algorithm in which the velocity is advected by the nonlinear term prior to being projected into a divergence-free space by $\mathbb{P}$.

While viscosity is neglected here, the difficulties encountered in stabilizing the pressure projection method for the Navier-Stokes equations are all encountered here as well.  The presence of viscosity will only aid in damping the numerical instabilities driven by the nonlinear advection term, so stability in the inviscid case is more difficult to achieve due to the absence of physical viscous dissipation.  In fact, the discussion of stability and under-resolution is primarily manifest in advection-dominated Navier-Stokes simulations in which a broad range of scales are present due to the lack of strong viscosity; in this sense then, the incompressible Euler simulations presented here capture the essence of the difficulty in simulating incompressible fully-viscous flows.

The numerical method used to model the density-stratified inviscid incompressible Euler equations is the spectral multi-domain penalty method (SMPM), a high-order discontinuous variant of the spectral element method \cite{Hesthaven1998, Escobar-Vargas2014} which has been previously shown to be effective in simulating high-Reynolds number environmental flows \cite{Diamessis2006,Diamessis2011} using the pressure-projection method.    In particular, we will use as an example the inviscid propagation of a solitary wave in a density-stratified channel as a test bed for evaluating the efficacy of the various post-processing methods.    In these simulations, the initial conditions propagate as waves in a non-dispersive non-dissipative fashion through the domain while retaining their form.  Thus, the degree to which these solutions maintain their structure is a good heuristic for the efficacy of these post-processing methods.     
 
The paper is organized as follows.  In Section 2 the exact and weak null-space projection methods, their motivation and the notation used are described.  In Section 3 two simulations with each of the three methods are conducted and compared.  Both simulations are of the same propagating solitary wave in tank, and the simulations differ in their mesh resolution.  In Section 4 is a discussion of the results presented in Section 3, along with a computational assessment of the spectrum and numerical conditioning properties of all of the methods compared in this paper.  Finally, we conclude with a short discussion of applicability to other numerical methods as well as a discussion of future work related to the ideas outlined herein. 

\section{Methods}

This section summarizes the the exact null-space projection (ENP) as outlined in Ref.~\cite{Steinmoeller2013}, the weak null-space projection which is the contribution of this work, and the numerical method.  

\subsection{Numerical method and notation}
\label{notation}

In the 2D SMPM, each element is assumed to be smoothly and invertibly mapped from the unit square $[-1,1]\times[-1,1]$ and the element connectivity is logically cartesian (each element has a single neighbor in each of the North, South, East, and West directions).
Within each element lies a two-dimensional Gauss-Lobatto-Legendre (GLL) grid; denote as $n$ the number of GLL points per direction per element, and $m_x$ and $m_z$ the number of $x$ and $z$ elements in the grid\footnote{Here $z$ is the vertical direction as is convention in environmental fluid mechanics.}.   Thus, the total number of grid points is $r = n^2 m_x m_z$.  
On the GLL grid, a two-dimensional nodal Lagrange interpolant basis of polynomial order $n+1$ is constructed such that each basis function has unit value on one of the $n^2$ GLL points and zero on all of the others.
This nodal basis is used for approximating functions and their derivatives which are calculated by way of spectral differentiation matrices \cite{Costa2000} which compute derivatives of nodally-represented functions by multiplying the nodal values by derivatives of the Lagrange interpolants themselves.
The SMPM is a discontinuous method and so $C^0/C^1$ inter-element continuity and boundary conditions are only weakly enforced.

On this grid, denote the differentiation operators $D_x, D_z \in \R^{r \times r}$.  As unsymmetric matrices that operate element-by-element, each is permutation-equivalent to a block-diagonal matrix.  The operations of gradient and divergence are defined and denoted as
\begin{align}
G &= \left[ \begin{array}{ccc}
D_x \\
D_z \\ 
\end{array} \right] \in \R^{2r \times r} \\
D &= [ D_x,  D_z] \in \R^{r \times 2r}
\end{align}
and it is emphasized that unlike in symmetric discretizations, $G \neq D^T$.   Finally we denote the Laplacian as $L \in \R^{r\times r}$.  With this notation, the discrete pressure projection operator $\mathbb{P} \in \R^{2r \times 2r}$ is defined
\begin{align}
	\mathbb{P} = I - G L^{-1} D.
\end{align}

Finally we denote as $\mathcal{N}$ the nonlinear term in the momentum equations
\begin{align}
	\mathcal{N}(\bf{u}) =  \bf{u} \cdot \nabla \bf{u} - g \frac{\rho'}{{\rho_0}}\bf{e}_z.
\end{align}

\subsection{No post-processing}
\label{no_post_processing}

Using first-order forward Euler time-stepping, a single time-step of the pressure-projection algorithm is given in Algorithm \ref{alg_pp}, where $\Delta t$ denotes the time-step.  In practice an adaptive first-order forward Euler time-stepping method was employed, but for simplicity we do not include the adaptivity in the algorithmic summary.  Below, $u \in \R^{2r}$ and $\rho \in \R^{r}$ are vectors defined on the grid, and all operators are discretized as noted in subsection \ref{notation}.
	\begin{algorithm}[H]
		\begin{algorithmic}[1]	
			\REQUIRE $u,\rho$ at time $t$. 
			\STATE $u^* = u + \Delta t \mathcal{N}(u)$
			\STATE $\rho \longleftarrow \rho  - \Delta t D u (\rho' + \overline{\rho} )$
			\STATE $u \longleftarrow \mathbb{P}u^*$
			\RETURN $u,\rho$ at time $t + \Delta t$.
		\end{algorithmic}
		\caption{A single time-step of the pressure projection method with no post-processing.}
		\label{alg_pp}
	\end{algorithm}

\subsubsection{The projection method in a discontinuous discretization} 	

\label{failure_of_pp}
Recalling that $\mathbb{P} = I - GL^{-1}D$ where $L$ is the Poisson operator, it is worth pausing to discuss why this projection method fails to yield an exactly divergence-free velocity field in the discrete case.  Consider the definition of $L \in \R^{r \times r}$ as a matrix.  $L$ certainly contains the product of the divergence with the gradient, $DG \in \R^{r\times r}$, but must also contain terms related to inter-element continuity conditions and boundary conditions.  Denoting these additional terms as $C \in \R^{r \times r}$, write $L$ as 
\begin{align}
	L = DG + C.
\end{align}
In the case of the spectral multi-domain penalty method $C$ represents the penalty conditions that enforce inter-element continuity and boundary conditions \cite{Escobar-Vargas2014}; in the case of discontinuous Galerkin $C$ represents the inter-element fluxes in the DG residual \cite{Steinmoeller2013}.  Now consider the divergence of the updated velocity, 
\begin{align}
	D\mathbb{P}u &= D (  I - GL^{-1}D ) u \nonumber \\
				&= Du - D G L^{-1} Du. \label{divcondition}
\end{align}
If $L = DG$, then it is clear that $D\mathbb{P}u = 0$, but, as already stated, $L$ contains other components encapsulated in $C$, and thus Eq.~(\ref{divcondition}) imposes conditions on $C$ to ensure that $D\mathbb{P}u = 0$.  Denoting as $v = L^{-1}Du$ the solution of the Laplace problem, note that $v$ satisfies $Lv = Du$ and so
\begin{align}
	DGv = Du - Cv.
\end{align}
Substituting this expression into Eq.~(\ref{divcondition}) for $L^{-1}Du$, we obtain 
\begin{align}
D\mathbb{P}u &= D (  I - GL^{-1}D ) u\nonumber \\
			&= Du - ( Du - Cv ) \nonumber  \\
			&= Cv.
\end{align}
This shows that the divergence of the projected velocity is equal to the discontinuity (as measured by $C$) in the solution $v \in \R^{r}$ of the Laplace problem $Lv = Du$.  As discussed above, the inter-element continuity conditions embedded within $C$ may be of flux-type or  $C^0/C^1$ type depending on the discretization method being employed. Nevertheless, assuming a discontinuous discretization, in general it is not true that $Cv = 0$, thus in general the projection method outlined in Algorithm \ref{alg_pp} does not yield an exactly divergence-free velocity; the divergence of $u$ is equal to the discontinuity of the Poisson solution $v$.  Finally it should be noted that although often the culprit of non-solenoidal vector fields and thus instability is the aliasing effects within the nonlinear advection term, this analysis  (corroborating that in Ref.~\cite{Steinmoeller2013}) shows that there is another source of spurious divergence embedded within the pressure projection operator for discontinuous element-based methods.


\subsection{Exact null-space projection}

To address the fact that the divergence-free condition is not satisfied (as described in Section \ref{failure_of_pp}), Steinmoeller et al. \cite{Steinmoeller2013} construct a post-processing method that is now summarized in this section.  Noting that as a matrix $D$ is element-local, computing the null space basis of $D$ is trivially parallelizable.  A local SVD of each block of $D$ can be computed in parallel and the right singular vectors corresponding to zero singular values can be retained.  Assuming that $D$ has a right nullity of dimension $k$, denote as $N \in \R^{2r \times k}$ the matrix whose columns are these null-value singular vectors.  $N$ as a matrix is orthonormal and $\norm{DN }= 0$.  The columns of $N$ are exactly all of the solenoidal vectors supported on the grid, and so, as in Ref.~\cite{Steinmoeller2013}, these vectors are used to project the velocity $u$ into the column space of $N$ by

\begin{align}
	u = NN^T \mathbb{P}u^*.
\end{align}

Denoting the composition of the pressure projection with the exact null-space projection as  $\mathbb{P}_e = NN^T \mathbb{P}$, observe that by construction $\mathbb{P}_e u$ is exactly divergence-free for any velocity $u\in\R^{2r}$ since $\norm{DNx} = 0$ for all vectors $x \in \R^{2n}$.  This algorithm, which we call \emph{exact null-space projection}, is summarized in Algorithm \ref{alg_enp}. 

	\begin{algorithm}[H]
		\begin{algorithmic}[1]	
			\REQUIRE $u,\rho$ at time $t$. 
			\STATE $u^* = u + \Delta t \mathcal{N}(u)$
			\STATE $\rho \longleftarrow \rho  - \Delta t D u (\rho' + \overline{\rho} )$
			\STATE $u \longleftarrow\mathbb{P}_eu^*$
			\RETURN $u,\rho$ at time $t + \Delta t$.
		\end{algorithmic}
		\caption{A single time-step of the pressure projection method with exact null-space projection.}
		\label{alg_enp}
	\end{algorithm}
Finally, it is worth noting that $\mathbb{P}_e$ is a projection in its own right since it is a composition of two projections.  

\subsection{Weak null-space projection}

While on one hand, the element-local nature of the divergence operator $D$ allows for the easy computation of its null-space basis $N$, this element-local nature also means that $NN^T$ as a projection may introduce discontinuities at the element interfaces.  Generally speaking, such discontinuities lead to steep gradients at the element interfaces and consequently instability. Motivated by this intuition, we now describe a second method of post processing the pressure-projection operator $\mathbb{P}$ which we call \emph{weak null-space projection} (WNP).

First, consider the following minimization problem in which $k$ is the column dimension of $N$, and $u \in \R^{2r}$ is a velocity vector: 
\begin{align}
	\lambda' =	\argmin_{\lambda \in \R^{k}} \norm{N\lambda - u}^2.
\end{align}
This least squares problem seeks a divergence-free velocity vector $N\lambda$ that minimizes the distance to $u$.  Of course the solution to this problem is $\lambda' = N^Tu$ and thus the divergence-free velocity is $u' = NN^Tu$, which is the ENP method outlined in the previous section.  

Treating the ENP as a least squares problem, we seek to penalize the minimization of the residual for discontinuity in the velocity.  Suppose there is an inter-element discontinuity operator $E \in \R^{2r \times 2r}$ that computes the inter-element discontinuity in a velocity $u \in \R^2r$.  For example in the discontinuous Galerkin approximation $E$ might be the numerical fluxes between elements, or in the SMPM it may be the inter-element penalty conditions.  In any case, to penalize both divergence and discontinuity in the residual, we write the following least squares problem:
\begin{align}
	\label{wnp_lsq}
 	u' = \argmin_{u' \in \R^{2r}} \norm{u' - u}^2 + \alpha_1 \norm{Du'}^2 + \alpha_2\norm{Eu'}^2.
 \end{align}
The solution of this least squares problem is given by the solution $u'$ to its normal equations,
\begin{align}
		(I + \alpha_1 D^TD + \alpha_2 E^TE)u' = u,
\end{align}
and a unique solution exists for all $\alpha_1, \alpha_2 > 0$, since the normal equations are then a symmetric, positive definite linear system and thus invertible.  Composing the solution to the normal equations with the pressure-projection, we denote the modified projection operator as 
\begin{align}
	\mathbb{P}_w = (I + \alpha_1 D^TD + \alpha_2 E^TE)^{-1} \mathbb{P}
\end{align} 
and note the algorithm for one time-step of this method is summarized in Algorithm \ref{alg_wnp}. 
	\begin{algorithm}[H]
		\begin{algorithmic}[1]	
			\REQUIRE $u,\rho$ at time $t$. 
			\STATE $u^* = u + \Delta t \mathcal{N}(u)$
			\STATE $\rho \longleftarrow \rho  - \Delta t D u (\rho' + \overline{\rho} )$
			\STATE $u \longleftarrow\mathbb{P}_wu^*$
			\RETURN $u,\rho$ at time $t + \Delta t$.
		\end{algorithmic}
		\caption{A single time-step of the pressure projection method with weak null-space projection.}
		\label{alg_wnp}
	\end{algorithm}
	
Notice a few things about this method.  First, if $\alpha_1 = \alpha_2 = 0$ then $\mathbb{P}_w = \mathbb{P}$ in which case no post-processing is being done and Algorithm \ref{alg_wnp} is identical to Algorithm \ref{alg_pp}.  Second, notice that $\mathbb{P}_w$ is not a projection in the linear algebraic sense since $\mathbb{P}_w ^2 \neq \mathbb{P}_w$.  However, notice that all eigenvalues of $(I + \alpha_1 D^TD + \alpha_2 E^TE)^{-1}$ lie within $[0,1]$; since projections also satisfy this property (although they \emph{only} have zero and one as eigenvalues) we claim that $\mathbb{P}_w$ is similar to a projection.  Finally, note that the solution of the regularized least squares problem defined in Eq.~(\ref{wnp_lsq}) will \emph{not} be exactly divergence-free, but the degree to which the divergence-free condition is satisfied can be controlled by the magnitude of $\alpha_1$, with a similar argument holding for continuity and $\alpha_2$.  

One may well ask why an exact projection onto the intersection of the spaces of divergence-free and continuous functions was not constructed.  This is primarily because the construction of an orthogonal basis for the space of continuous functions is impractical, as it requires the factorization of matrices of dimension $r$ that are not element-local (e.g. obtaining the right null space basis of $E$).  Moreover, it is not clear that such a projection would even be desired, given the suggested numerical benefits of allowing a discontinuous discretization \cite{Diamessis2005, Restelli2009, Hesthaven1998}. 

Notice also that the weak null-space projection method is a generalization, in a sense, of both the pressure-projection and exact null-space projection.  If $\alpha_1 = \alpha_2 = 0$ then there is no post-processing being done and this method is identical to Algorithm \ref{alg_pp}.  If $\alpha_2 = 0$ then this method emphasizes minimizing divergence at the cost of continuity, and its solution approaches that of Algorithm \ref{alg_enp} in the limit that $\alpha_2 \gg 1$\footnote{This can be proven by considering the eigendecomposition of the normal equations and taking the limit as $\alpha_2 \Rightarrow \infty$}.  Thus, weak null-space projection is a generalization of both of the other two methods which appear as special cases of it. 

Finally note that the idea of solving a least squares problem as a means of enforcing the inter-element continuity and boundary conditions within a Navier-Stokes simulation has been explored previously in the case of least-squares spectral element methods \cite{Proot2005}, but these methods suffer from the ill-conditioning of the normal equations of already ill-conditioned element matrices.  By using these least-squares ideas as a post-processing method alone, we can somewhat control the ill-conditioning of the normal equations, as will be discussed in Section \ref{sec_condition}.

\section{Results}

In this section, the three algorithms for post-processing are compared in two simulations of a propagating solitary wave in a density-stratified tank of water.  The three methods compared are henceforth called \emph{pressure projection}, \emph{exact null-space projection}, and \emph{weak null-space projection methods} and correspond respectively to Algorithms \ref{alg_pp}, \ref{alg_enp}, and \ref{alg_wnp}.  The simulations model a propagating solitary wave and are initialized by an eigenfunction solution of the Dubreil-Jacotin-Long (DJL) equation \cite{Dunphy2011}.  These waves are exact solitary solutions of the Euler-Boussinesq equations meaning that through a balance of nonlinearity and dissipation, they preserve their shape as they propagate through the domain.  Thus, a simulation ought to explicitly preserve the waveform as it propagates, and so this initial condition provides a convenient test of the accuracy of a numerical method.

 The computational domain represents a fluid-filled tank and is a rectangle of width $12$ m and height $0.15$m.  The wave has a phase speed of $c = 0.104$ m/s, and a wavelength of about $1.5$ meters.  The wave is allowed to propagate for approximately $60$ seconds during which time it travels about four wavelengths.  Stability of these methods is assessed by examining the duration for which the norm of the velocity remains bounded.  Two simulations of this physical problem are considered; in Section \ref{sec_hires} the mesh is chosen to be of high-enough resolution to clearly resolve the solitary wave.  In Section \ref{sec_lores} the same simulation is repeated but with a lower-resolution mesh that only marginally resolves the wave scale.  All simulations use a simple adaptive first-order Euler time-stepping in which is time-step in adjusted to maintain a CFL number of less than $0.1$.  This restrictive bound on the CFL condition is due to the low-order time-integration method used, but the focus of this study is divergence-free enforcement in the projection method and not the time-stepping method, so the time-stepping method is kept simple for ease of implementation.

\subsection{Well-resolved simulation}

\label{sec_hires}
	\begin{figure}[!h]
		\begin{center}
		\includegraphics[width=0.95\textwidth]{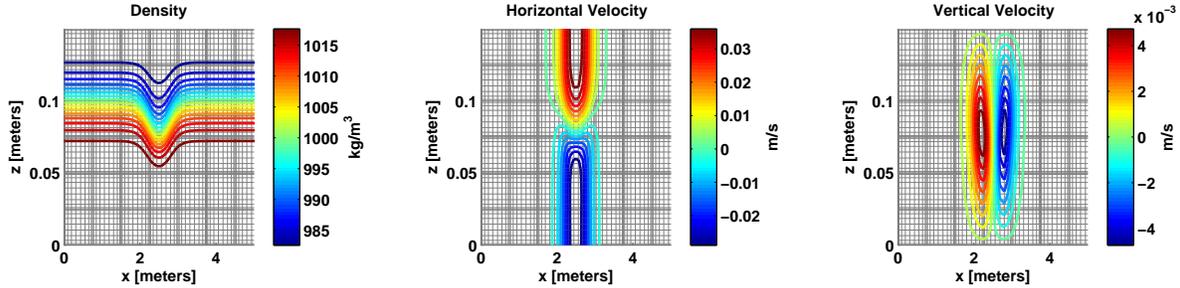}
		\end{center}
		\caption{The initial conditions of the well-resolved simulation in which $n = 10, m_x = 12, m_z = 6$, with only the first $5$ meters of a $12$-meter-long domain shown.}
		\label{fig_ic_hires}
	\end{figure}
	
The well resolved simulation uses a mesh with $m_x = 16$ elements in the horizontal direction and $m_z = 6$ elements in the vertical direction.  Each element has $n = 10$ grid points per direction, resulting in $100$ Gauss-Lobatto-Legendre points per element and a total of $r = 9600$ grid points.  The initial condition and a snapshot of the first $5$ meters of the mesh are depicted in Fig.~\ref{fig_ic_hires}.  The wave is propagating to the right.  
		
	\begin{figure}[!h]
		\begin{center}
		\includegraphics[width=0.95\textwidth]{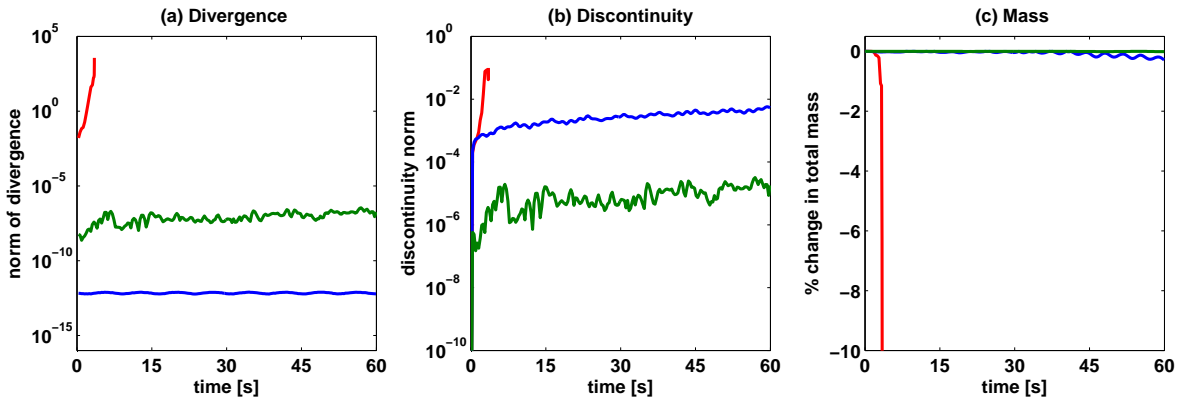}
		\end{center}
		\caption{The divergence norm, discontinuity norm, and total mass of the well-resolved simulation as a function of time.  Algorithm \ref{alg_pp} is shown in red, Algorithm \ref{alg_enp} in blue, and Algorithm \ref{alg_wnp} in green.  All of the data in the above plots are shown beginning from the second time-step onwards, as the first time-step has zero divergence and zero discontinuity for all methods. }
		\label{fig_hires_summary}
	\end{figure}	
	
	The results of this simulation are summarized in Fig.~\ref{fig_hires_summary} in four different ways.  Algorithm \ref{alg_pp} is shown in red, Algorithm \ref{alg_enp} in blue, and Algorithm \ref{alg_wnp} in green, and each of the time-series plots is shown in simulation window of $t \in [0,60]$ seconds or until the time when the simulation became unstable.  Fig.~\ref{fig_hires_summary}(a) depicts the norm divergence of the flow at each time-step.  This is calculated by taking the discrete 2-norm of the divergence of the discrete velocity vector and normalizing by the norm of the velocity, and is defined as
	\begin{align}
		\frac{\norm{Du}}{\norm{u}}(t) = \frac{\norm{D_x u_x(t) + D_z u_z(t)}}{\sqrt{u_x(t)^2 + u_z(t)^2}}
	\end{align}
	and as such is a  scalar function of time.  The results in Fig.~\ref{fig_hires_summary}(a) demonstrate that the rapid growth of divergence in the pressure-projection simulation (red) drives instability, leading to blowup at about time $t =5$ seconds.  As postulated in Section \ref{failure_of_pp}, the divergence grows quickly at the inter-element interfaces and eventually spreads into the interior of the elements.  By contrast, both the post-processed simulations (green and blue) remain stable throughout the simulation window, though it is clear that they reduce divergence to differing degrees -- the exact null-space projection reduces divergence near machine precision (blue curve), while the weak null-space projection (green curve) reduces the divergence to about $\mathcal{O}(10^{-10})$; tolerably small but not machine precision.  Nevertheless, since both post-processing algorithms are capable of maintaining stability by reducing the divergence of the flow, it is clear that divergence is the root of the instability.  Namely, discontinuities in the solution of the Poisson problem (c.f. Section \ref{failure_of_pp}) lead to spurious divergence at the element interfaces which in turn feeds into the nonlinear terms in the momentum equation which amplifies these numerical artifacts until the simulation becomes unstable.
	
	Fig.~\ref{fig_hires_summary}(b) depicts the degree to which the velocity field is continuous across element boundaries.  Discontinuity in this context is defined as
	\begin{align}
		\frac{\norm{Eu}}{\norm{u}}(t) = \frac{\norm{E[u_x; u_z]}}{\sqrt{u_x(t)^2 + u_z(t)^2}}
	\end{align} 
	where $E\in\R^{2r\times2r}$ is the inter-element continuity misfit functional as defined in the SMPM, which includes both $C^0$ and $C^1$ continuity conditions \cite{Escobar-Vargas2014}.  The rapid growth of discontinuity in the pressure-projection simulation (red) is just due to the instability; as $\norm{u}$ grows, naturally $\norm{Eu}/\norm{u}$ grows at a rate proportional to the operator norm of $E$.  The exact null-space projected velocity (blue) is discontinuous at about three orders of magnitude below the norm of the velocity, and appears to grow slowly over time.  This is still relatively continuous, but the growth in time suggests that over time the accumulated discontinuity may need to be addressed.  Finally, the weak null-space projected velocity (green) is the most continuous; the discontinuity norm is about five orders of magnitude smaller than the norm of the velocity.  Here too the continuity appears to worsen over time.
		
	Finally we measure the conservation properties of these methods by examining the percent change in mass (Fig.~\ref{fig_hires_summary}(c)) over time.  Assuming a unit transverse direction, the mass is computed with Gauss-Lobatto-Legendre quadrature-discretized version of the following integral:
	\begin{align}
		\textrm{total perturbed mass} (t) &= \sum_k\int_{\Omega_k} \rho'(x,z,t) d\Omega = \sum_{k}\sum_{ij} w_i w_j \rho_k(x_i,z_j,t) 
	\end{align}
	where $\Omega_k$ is an element in the mesh and  $w_j$ are the Gauss-Lobatto-Legendre quadrature weights.
	\begin{figure}[h]
		\begin{center}
		\includegraphics[width=0.95\textwidth]{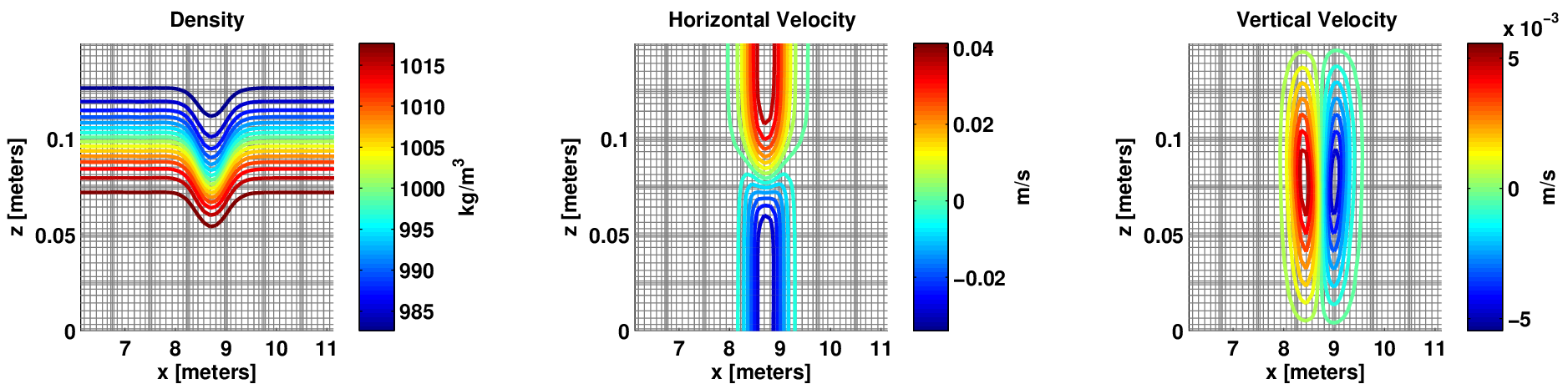}
		\end{center}
		\caption{The exact null-space projection method at time $t = 60$s in the well-resolved simulation in which $n = 10, m_x = 12, m_z = 6$; only $5$ meters of the $12$ meter-long domain are shown.}
		\label{fig_exact_final_hires}
	\end{figure}
	
\begin{figure}[h]
		\begin{center}
		\includegraphics[width=0.95\textwidth]{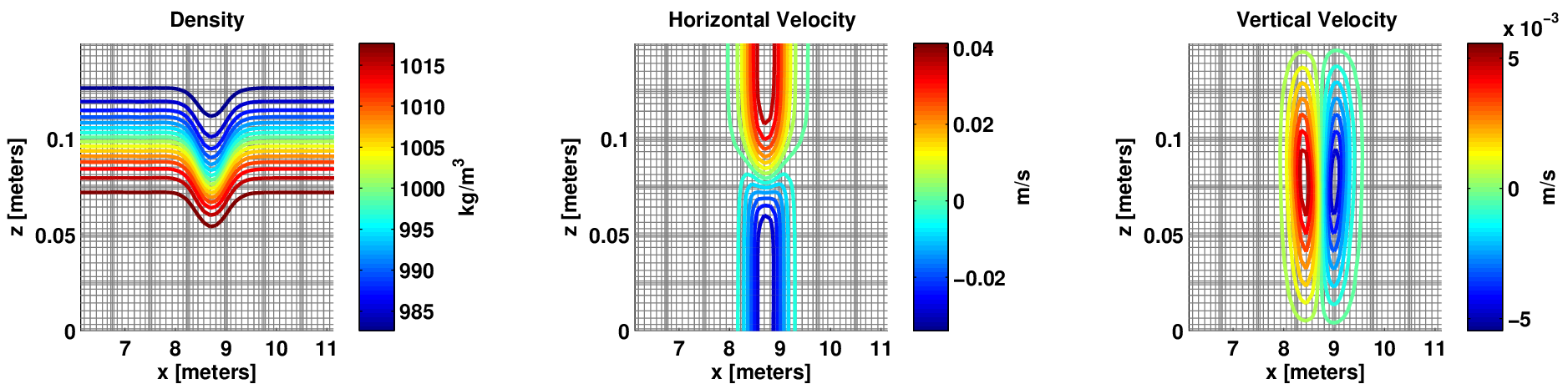}
		\end{center}
		\caption{The weak null-space projection method at time $t = 60$s in the well-resolved simulation in which $n = 10, m_x = 12, m_z = 6$; only $5$ meters of the $12$ meter-long domain are shown.}
		\label{fig_weak_final_hires}
	\end{figure}

	The change in mass is shown as a percentage in Figs.~\ref{fig_hires_summary}(c).  Both the exact (blue) and weak (red) null-space projections conserve mass relatively well.  The exact null-space projection is slightly less conservative losing about $0.25\%$ of total mass; the weak projection conserves mass to 5 decimal places.  	Of note is that both post-processing methods evolve the wave at the correct phase speed as determined by the DJL computation.  Finally, the wave at the final time of $t = 60$ seconds as computed by the exact null-space projection and weak null-space projection methods is depicted in Figures \ref{fig_exact_final_hires} and \ref{fig_weak_final_hires}.  The two solutions are visually indistinguishable and differ less than $1\%$ numerically, a discrepancy likely due to the slightly different final physical time that results from the adaptive time-stepping. 			
				
\subsection{Marginally-resolved simulation}

\label{sec_lores}
	\begin{figure}[!h]
		\begin{center}
		\includegraphics[width=0.95\textwidth]{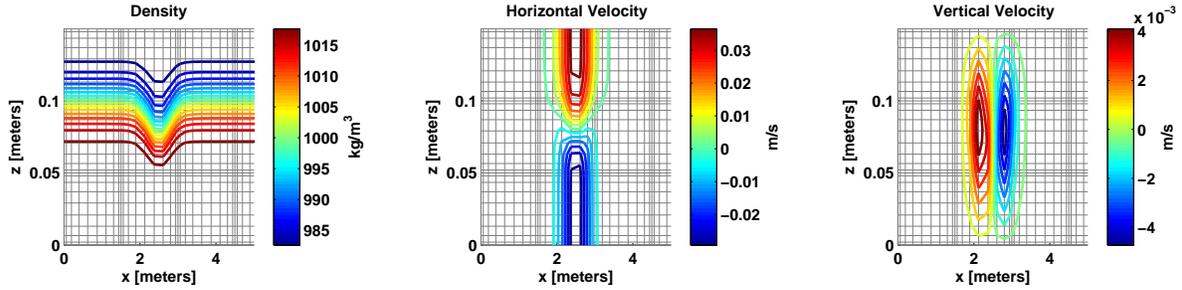}
		\end{center}
		\caption{The initial conditions of the marginally-resolved simulation in which $n = 10, m_x = 6, m_z = 3$; only the first $5$ meters of the $12$ meter-long domain are shown.}
		\label{fig_ic_lores}
	\end{figure}
	
\begin{figure}[!h]
		\begin{center}
		\includegraphics[width=0.95\textwidth]{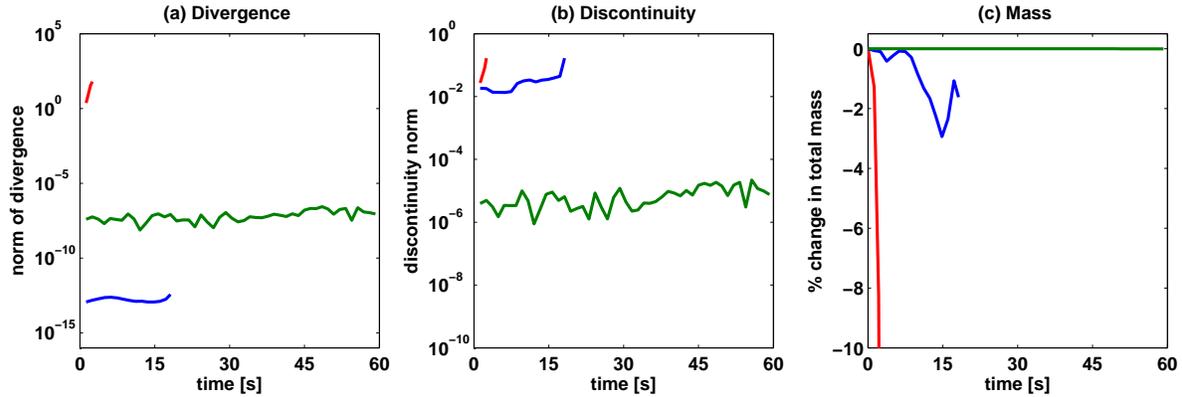}
		\end{center}
		\caption{The divergence norm, discontinuity norm, and the total mass of the marginally-resolved simulation as a function of time.  Algorithm \ref{alg_pp} is shown in red, Algorithm \ref{alg_enp} in blue, and Algorithm \ref{alg_wnp} in green.  All of the data in the above plots are shown beginning from the second time-step onwards, as the first time-step has zero divergence and zero discontinuity for all methods.}
		\label{fig_lores_summary}
	\end{figure}

The marginally-resolved simulation uses a mesh with one-quarter of the resolution of the previous mesh.  This coarse mesh has $m_x = 8$ elements in the horizontal direction and $m_z = 3$ elements in the vertical direction.  Each element has $n = 10$ grid points per direction, resulting in $100$ Gauss-Lobatto-Legendre points per element and a total of $r = 2400$ grid points.   The initial condition and a snapshot of the first $5$ meters of this mesh are depicted in Fig.~\ref{fig_ic_lores};  again the wave propagates to the right.
As before, the results of this simulation are summarized in Fig.~\ref{fig_lores_summary} in four different ways and Algorithm \ref{alg_pp} is shown in red, Algorithm \ref{alg_enp} in blue, and Algorithm \ref{alg_wnp} in green.  Each of the time-series plots is shown in the simulation time window of $t \in [0,60]$ seconds or until the simulation became unstable.   	
	
Notice first in Fig.~\ref{fig_lores_summary}(a) that although the ENP method (blue) is divergence-free throughout the time that it is stable, it is only stable until time $t = 16$ seconds.  Conversely, the weak null-space projection (WNP) method remains stable throughout the simulation window of $60$ seconds, even though it is several orders of magnitude more divergent.  A possible explanation of this is found in Fig.~\ref{fig_lores_summary}(b), which shows that the ENP method is three orders of magnitude more discontinuous than the WNP method, and the discontinuity grows rapidly just before the simulation becomes unstable.  Examining Fig.~\ref{fig_lores_summary}(c), it is also clear that the ENP does not conserve mass (as the WNP does) prior to becoming unstable.  Finally note that the velocity fields exhibit high-frequency oscillations at element interfaces that are likely due to a lack of filtering or de-aliasing.  
	
	\begin{figure}[h]
		\begin{center}
		\includegraphics[width=0.95\textwidth]{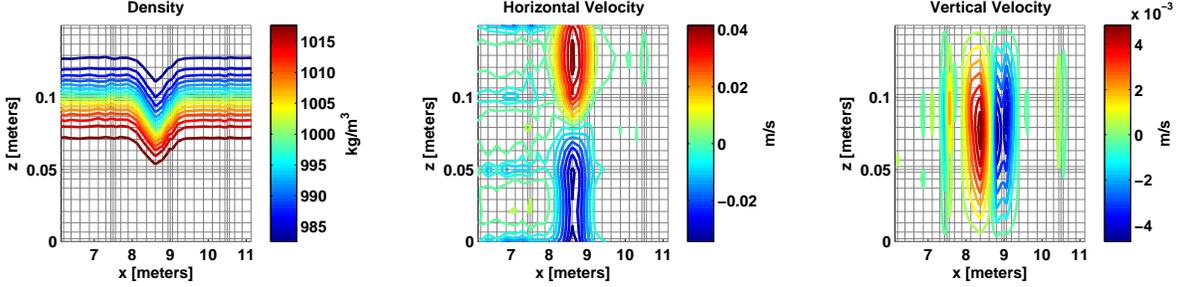}
		\end{center}
		\caption{The weak null-space projection method at time $t = 60$s in the marginally-resolved simulation in which $n = 10, m_x = 12, m_z = 6$; only $5$ meters of the $12$ meter-long domain are shown..}
		\label{fig_weak_final_lores}
	\end{figure}
	
\section{Discussion}

\subsection{Conditioning of the weak projection operator}
\label{sec_condition}
The 2-norm condition number of a symmetric positive definite matrix $A$ is defined as the ratio of its maximum to its minimum eigenvalue,
\begin{align}
	\kappa_2 (A) =  \frac{  \lambda_{\textrm{max}}(A) }{ \lambda_{\textrm{min}}(A) }.
\end{align}
The normal equations have the form $ I + \alpha( D^TD + E^TE )$, which means that their condition number as matrix is
\begin{align}
\kappa_2 (I + \alpha D^TD + \alpha E^TE)  =  \frac{ 1 + \alpha  \lambda_{\textrm{max}}(D^TD + E^TE) }{1 + \alpha  \lambda_{\textrm{min}}(D^TD + E^TE) } .
\end{align}
Since there exist velocity vectors that are both divergence-free and continuous (for example, any constant velocity), we know that there is a vector $v \in \R^{2r}$ such that $(D^TD + E^TE)v = 0$, which means that $\lambda_{\textrm{min}}(D^TD + E^TE) = 0$.  Thus the condition number of the normal equations is
\begin{align}
\kappa_2 (I + \alpha D^TD + \alpha E^TE)  =1 + \alpha  \lambda_{\textrm{max}}(D^TD + E^TE).
\end{align}
This means that the conditioning of the normal equations scales linearly with $\alpha$, and so the greater the degree to which we penalize discontinuity  and divergence, the more difficult and less accurate the solution of the linear system becomes.

\subsection{Study of various values of the regularization coefficients}

	\begin{figure}[!h]
		\begin{center}
		\includegraphics[width=0.95\textwidth]{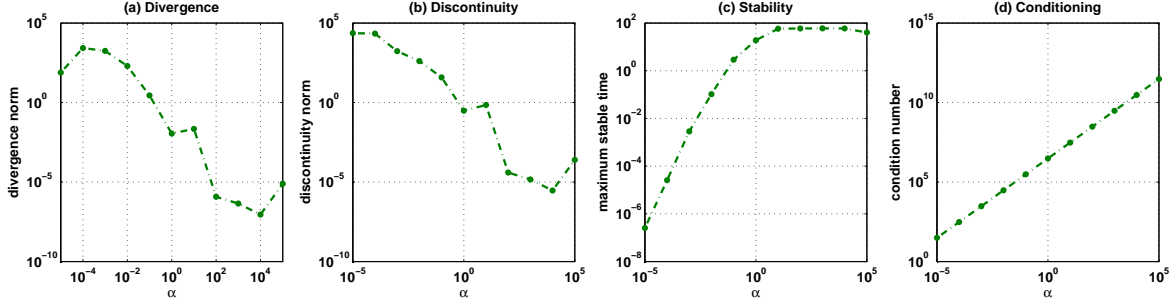}
		\end{center}
		\caption{The divergence norm, discontinuity norm, maximum stable time, and condition number of the marginally-resolved simulation as a function of $\alpha = \alpha_1 = \alpha_2$.}
		\label{fig_alpha}
	\end{figure}
	
Finding an optimal choice of the regularization coefficients $\alpha_1$ and $\alpha_2$ is in general a difficult problem, especially in this context in which optimality is really determined by stability.  To obtain an insight into the general performance of the weak projection method as a function of $\alpha = \alpha_1 = \alpha_2$, the low-resolution problem (Section \ref{sec_lores}) was solved for values of $\alpha = [10^{-5}, 10^{-4}, \dots, 10^{5}]$.   The results of this set of simulations are summarized in Fig.~\ref{fig_alpha}, in which the divergence norm, the discontinuity norm, the maximum stable time, and the condition number are all plotted as a function of $\alpha$ at the final time the simulation remained stable.  As $\alpha$ grows, generally the divergence (Fig.~\ref{fig_alpha}(a)) and discontinuity (Fig.~\ref{fig_alpha}(b)) in the simulation decrease, but the condition number (Fig.~\ref{fig_alpha}(d)) grows linearly (see discussion in Section \ref{sec_condition}).  This results in increased difficulty in solving the normal equations and eventually a loss of accuracy that manifests as an increase in divergence and discontinuity (as observed in the $\alpha > 10^4$ regime in Figs.~\ref{fig_alpha}(a) and (b)).  For all $\alpha > 10^{0}$ the simulation is stable throughout the simulation window of $60$ seconds.  While this analysis is not conclusive, it suggests that an optimal choice of $\alpha$ exists, one that minimizes the condition number while still ensuring a divergence-free and continuous velocity field that remains stable.  
	
\subsection{Spectral analysis of pressure update operators}
\label{sec_spectral}
	\begin{figure}[!h]
		\begin{center}
		\includegraphics[width=0.75\textwidth]{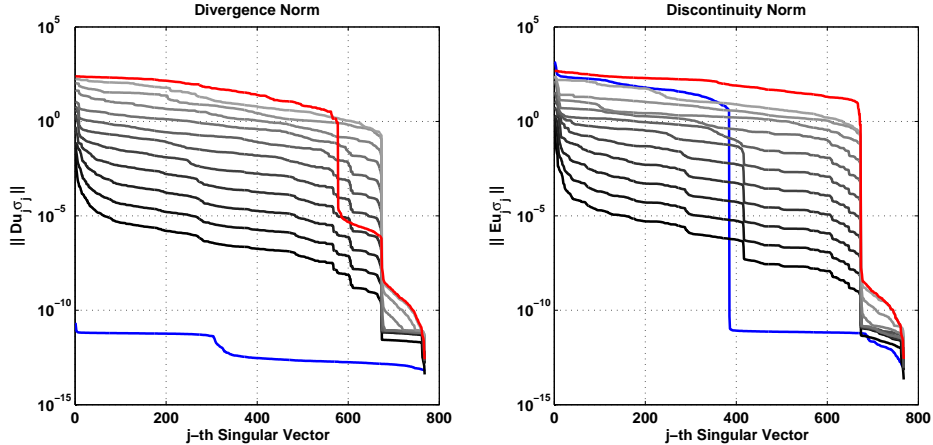}
		\end{center}
		\caption{The divergence (a) and discontinuity (b) of the left singular vectors that span the range space of each pressure update operator as a function of $\alpha$.  In blue is the exact null space projection and in red is pressure projection with no post-processing.  Darker colors mean larger values of $\alpha$ with $\alpha \in [10^{-5}, 10^{5}]$.  The singular vectors were sorted by their divergence norm (left) and discontinuity norm (right), respectively. }
		\label{fig_spectrum}
	\end{figure}

While it is clear that the weak null-space formulation is effective at reducing discontinuity and divergence, it is instructive to observe exactly how the range space of the pressure update operator is modified as $\alpha = \alpha_1 = \alpha_2$ is changed.  Recall that given a matrix $A\in\R^{n \times n}$ the singular value decomposition (SVD) is given as
\begin{align}
	A = USV^T
\end{align}
where $U,V \in \R^{n \times n }$ are unitary and $S \in \R^{n\times n}$ is a diagonal matrix.  The range space of $A$ is given by the singular vectors (e.g. columns of $U$) that correspond to non-zero singular values of $A$ (e.g. the diagonal entries of $S$).  It is the space of all possible vectors $Ax \in \R^n$ where $x \in \R^n$, and so in the context of the velocity update operators in the projection method the range space represents all possible updated velocities.  To study the divergence and the continuity properties of the range space of the projection methods, then, we compute the SVD of $\mathbb{P}$, $\mathbb{P}_e$, and $\mathbb{P}_w(\alpha)$ for various values of $\alpha$.  Then for each left singular vector $u_j$ and corresponding singular value $\sigma_j$ we can compute its divergence
\begin{align}
	\norm{\sigma_j Du_j }
\end{align}
and discontinuity
\begin{align}
	\norm{\sigma_j E u_j }.
\end{align}
The magnitude of these values characterize the degree to which each singular vector is divergent and discontinuous, and the set of these values help characterize the divergence and discontinuity of the spectrum and range space of each of our pressure update operators $\mathbb{P}$, $\mathbb{P}_e$, and $\mathbb{P}_w(\alpha)$.

This computation is shown in Fig.~\ref{fig_spectrum}, with the divergence shown on the left and the discontinuity shown on the right.  Each curve depicts a descending sorted set of values $\{\sigma_j\norm{ Du_j}\}_{j=1}^{2r}$ or $\{\sigma_j\norm{ Eu_j}\}_{j=1}^{2r}$ plotted against $j$. The values for the pressure update operator with no post-projection are shown in red, and the values for the exact null-space projection are shown in blue.  In shades of gray are the values for the weak null-space projection, ranging for $\alpha = [10^{-5}, 10^{-4}, \dots, 10^{5}]$ with darker shades representing larger values of $\alpha$. 

First notice in the left panel of Fig.~\ref{fig_spectrum} that as $\alpha$ grows from $10^{-5}$ to $10^{5}$ more and more modes of $\mathbb{P}_w(\alpha)$ become divergence-free, and the rate of decay of lower (higher index $j$) modes decay faster.  This indicates, naturally, that as $\alpha$ grows, the range space of $\mathbb{P}_w(\alpha)$ becomes more and more solenoidal.  Second, notice that for all values of $\alpha$ the vast majority of modes of $\mathbb{P}_w(\alpha)$ are less divergent than those of $\mathbb{P}$; this is a good indication that the post-processing will reduce the divergence in  the velocity field.  Finally, notice that all modes of $\mathbb{P}_e$ are less divergent than any of the other modes of any of the other operators, as is expected. 

The right panel of Fig.~\ref{fig_spectrum} depicts the discontinuity of each of the singular vectors of the three operators $\mathbb{P}$, $\mathbb{P}_e$, and $\mathbb{P}_w(\alpha)$.  Notice first that first few modes of $\mathbb{P}_e$, shown in blue, are very discontinuous.  These modes are the dominant modes of the range space of $\mathbb{P}_e$, and so contribute significantly to its amplification of discontinuity.   Secondly, notice that the shape of $\mathbb{P}_w(\alpha)$ changes dramatically from $\alpha = 10^{-1}$ to $\alpha = 10^0$, exhibiting a steep drop in the discontinuity of the spectrum around the  $j = 400$ mode.  This transition is also the same place the stability properties reach a plateau as is evidenced in Fig.~\ref{fig_alpha}(c), indicating that discontinuity of the pressure projection operator and stability are at least well-correlated.  With regards to choosing a value of $\alpha$, it is observed that for $\alpha > 10^0$, the shape of the $\mathbb{P}_w(\alpha)$ curve doesn't change, but its overall value continues to decrease.  Along with the evidence shown in Fig.~\ref{fig_alpha}(c), this suggests that a fundamental property of the weak projection operator is unchanged after $\alpha > 1$ in which regime the method is stable.  And finally, as observed in practice, $\mathbb{P}_w$ exhibits significantly better continuity properties than either $\mathbb{P}$ and $\mathbb{P}_e$; this is not unexpected as the the latter two projection methods do not consider continuity at all in their post-processing.

\section{Summary and Conclusions}

\subsection{Summary}
In this paper, a method for post-processing the projection operation in the pressure projection method was developed that reduces inter-element discontinuity along with divergence as part of a time-evolving incompressible Euler simulation.  This method was compared with an exactly divergence-free projection method as developed in Ref.\cite{Steinmoeller2013} on a simulation of a propagating solitary wave.  While the exact null-space projection method (e.g. as in Ref.~\cite{Steinmoeller2013}) was better at mitigating divergence than the weak null-space projection, when the mesh resolution was coarsened so that the flow was only marginally resolved the weak null-space projection maintained stability for at least four times as long.  The continuity and conservation properties of the flow field were also observed to be better in the weak null-space projection method, and both weak and exact null-space projection methods outperformed the simulation without any post-processing.  

The drawback of the weak null-space projection method is that it requires the solution of a large, sparse, positive-definite linear system in solving the normal equations.  The difficulty of computing such a solution at scale was not addressed here, and is likely a significant computational challenge for an iterative method.  However, the benefits of this method, particularly the ability to remain stable in an under-resolved context (or conversely to reduce resolution while maintaining stability), suggest that this considerable computational effort is worth undertaking. 

\subsection{Extension}

The results described here are limited in the sense that they are only demonstrated on a single test case, that of a propagating solitary wave, and for a single discretization, the spectral multi domain penalty method.  However, the ideas described are rather general in that they apply to any discontinuous element-based discretization of the pressure projection method.  A case can be made that although the efficacy of the weak projection method needs to be demonstrated on a broader class of problems, the spectral analysis presented in Section \ref{sec_spectral} is independent of the problem being solved and demonstrates that the WNP method preserves continuity and reduces divergence simultaneously.   

\subsection{Future Work}

The natural extension is to ask whether $\alpha_1$ need be equal to $\alpha_2$, and whether either should vary in time.  In particular, if $\alpha_1(t)$ and $\alpha_2(t)$ are time-dependant, then it might be possible to choose them such that
	\begin{align}
		\alpha_1(t) &\propto \norm{Du}(t) \\
		\alpha_2(t) &\propto \norm{Eu}(t)
	\end{align}
	as a way to dynamically mitigate the effects of divergence and discontinuity.   So, if a flow were to being to exhibit troublesome discontinuities or divergence, then the coefficients could be dynamically adjusted to compensate and presumably improve the divergence and continuity properties of the flow.  It may even be possible to let $\alpha$ vary in space as well as time and become a weighting function for the residual that would locally adjust to values of the divergence and discontinuity.  In this way it may be possible to minimize the penalty paid in ill-conditioning for large values of $\alpha$ while still reaping the benefits of a smoother more divergence-free solution in places where it is critical.  
	
	Finally, and as mentioned previously in Section \ref{sec_lores}, there are significant numerical artifacts that are likely due to aliasing effects due to the nonlinear term.  If this is indeed due to the nonlinear term, then some sort of filtering or dealiasing is necessary to damp spurious high-frequency aliased components.  Worth investigating, then, is how these post-processing methods would interact with filtering or dealiasing.  

\section*{Acknowledgements}

The authors would also like to thank the United States Department of Defense High-Performance Computing Modernization Office for the NDSEG fellowship and the National Science Foundation for CAREER award \#0845558 for support.  Marek Stastna is supported by the Natural Sciences and Engineering Research Council of Canada.

\section*{References}

\bibliographystyle{elsarticle-harv}
\bibliography{bibliography}

\end{document}

%% file: macros.tex
\usepackage{amsmath}   
\usepackage{amsthm}
\usepackage{amssymb} 
\usepackage{calc}
\usepackage{ifthen}
\usepackage{graphicx}
\usepackage{enumerate}
\usepackage{centernot}
\usepackage[mathscr]{eucal}

\renewcommand\and{\ensuremath{\;\;\;\;\;\;\text{and}\;\;\;\;\;\;} }

\newcommand\restr[2]{{
  \left.\kern-\nulldelimiterspace 
  #1 
  \vphantom{\big|} 
  \right|_{#2} 
}}

 \newcommand\R{\ensuremath{\bb{R}}}

 \DeclareMathOperator*{\argmin}{arg\,min}
 
\newcommand\norm[1]{\ensuremath{\left\lvert \left\lvert  #1 \right\rvert \right\rvert }}

\newcommand\avg[2][1]{
	\ifthenelse{\equal{#1}{1}}{\ensuremath{\left \langle#2\right \rangle }}{}
	\ifthenelse{\equal{#1}{2}}{\ensuremath{\left \langle#2^2\right \rangle }}{}
	\ifthenelse{\equal{#1}{3}}{\ensuremath{\left \langle#2\right \rangle ^2}}{}}

\newcommand\derv[3][1]{
	\ifthenelse{\equal{#1}{1}}{\ensuremath{\frac{d#2}{d#3}}}{}
	\ifthenelse{\equal{#1}{2}}{\ensuremath{\frac{d^2#2}{d#3^2}}}{}}

\newcommand\prtl[3][1]{	\ifthenelse{\equal{#1}{1}}{\ensuremath{\frac{\partial#2}{\partial#3}}}{}
	\ifthenelse{\equal{#1}{2}}{\ensuremath{\frac{\partial^2#2}{\partial#3^2}}}{}}

\renewcommand\bf[1]{\ensuremath{\mathbf{#1}}}
\newcommand\bb[1]{\ensuremath{\mathbb{#1}}}

\renewcommand\eqref[1]{Eq. (\ref{#1})}

\theoremstyle{plain}   
\theoremstyle{definition}  
\numberwithin{exer}{subsection} 
\theoremstyle{plain}   
\theoremstyle{plain}   
\theoremstyle{definition}   
\theoremstyle{definition}   
\theoremstyle{plain}   
\theoremstyle{plain}   
\theoremstyle{plain}    
